\documentclass{acmart}

\usepackage{tabularx, booktabs, makecell, caption}
\usepackage{siunitx}
\usepackage{subcaption}
\usepackage{wrapfig}

\DeclareMathOperator*{\argmin}{arg\,min}

\acmConference[EEAMO '21]{ACM Conference on Equity and Access in Algorithms, Mechanisms, and Optimization}{October 5-9, 2021}{virtual}
\acmBooktitle{ACM Conference on Equity and Access in Algorithms, Mechanisms, and Optimization
October 5-9, 2021}
\acmPrice{}
\acmISBN{}

\renewcommand\footnotetextcopyrightpermission[1]{} 
\setcopyright{none}

\begin{document}

\title{Identifying biases in legal data: An algorithmic fairness perspective}


\author{Jackson Sargent}
\authornote{Work done while intern at Claudius Legal Intelligence.}
\email{jacsarge@umich.edu}
\affiliation{%
  \institution{University of Michigan}
   \country{United States}
}
\author{Melanie Weber}
\email{mw25@math.princeton.edu}
\authornote{Corresponding author.}
\affiliation{%
  \institution{Princeton University \& Claudius Legal Intelligence}
  \country{United States}
 }


\begin{abstract}
  The need to address \emph{representation biases} and \emph{sentencing disparities} in legal case data has long been recognized. Here, we study the problem of identifying and measuring biases in large-scale legal case data from an algorithmic fairness perspective. Our approach utilizes two regression models: A baseline that represents the decisions of a ``typical'' judge as given by the data and a ``fair'' judge that applies one of three fairness concepts. Comparing the decisions of the 
  ``typical'' judge and the ``fair'' judge allows for quantifying biases across demographic groups, as we demonstrate in four case studies on criminal data from Cook County (Illinois).
\end{abstract}


\maketitle

\section{Introduction}
A cornerstone of the development of trustworthy and discrimination-free machine learning and artificial intelligence is the consideration of biases in the training data. As artificial intelligence enters the legal space, it is essential to recognize biases in legal data and ensure that they are not replicated and reinforced with legal technology~\citep{Hao2019,Chouldechova2017,Madaio2020}. Furthermore, understanding biases in legal data and developing discrimination-free technology could help the legal space to become fairer and more widely accessible. 

We typically find two types of biases in legal data: First, \emph{representation biases}, i.e., certain social groups are over- or underrepresented in a data set. Second,  \emph{sentencing disparities}, i.e., the outcome of legal proceedings for similar cases varies across social groups.
Representation biases may reflect disparities in policing (arrest rates) or in offense rates. While potential differences in offense rates across demographic groups (e.g., male and female for specific crime types) are difficult to evaluate, policing disparities have been studied with data-driven methods through the analysis of arrest data.
The need to address
sentencing disparities has long been recognized. In the context of federal sentencing, the Sentencing Reform Act was passed in 1984 as part of the Comprehensive Crime Control Act with the aim of mitigating sentencing disparities. This has resulted in the establishment of the United States Sentencing Commission in 1987 to publicize federal sentencing guidelines. Today, both representation biases and sentencing disparities are active areas of legal scholarship~\citep{mustard2001, rehavi_starr,cohen_yang,Schanzenbach2007ReviewingTS,bias_law,avery2019racial,Kleinberg2018HumanDA}. 

Recently the data-driven analysis of representation biases and sentencing disparities has gained a surge of interest~\citep{rehavi_starr,Chouldechova2017,Berk2021}. The present study is guided by the following question:
\begin{quote}
\emph{How can we efficiently identify and measure biases in large-scale legal data?}
\end{quote}
Much of the existing body of literature focuses on the quantitative analysis of sentencing disparities~\citep{mustard2001, rehavi_starr,cohen_yang}, for instance, by analyzing correlations between the characteristics of the defendant -- and sometimes the judge -- with the case outcome, e.g., the length of the sentence. Here, we take a different approach.
We propose to compare the decisions of a ``fair'' judge with the judicial decisions in our data set. For this, we analyze two models: A baseline model trained via an unconstrained regression approach on the data (``average'' or ``typical'' judge) and a second model trained with \emph{fairness constraints} (``fair'' judge). 

In addition to studying sentencing disparities via algorithmic fairness metrics, we also analyze representation bias by comparing the performance of the baseline model with a second, balanced model. Finally, we compare with a balanced classifier that was trained with fairness constraints -- combining both bias mitigating approaches.

In contrast to much of the prior literature, which analyzes sentencing disparities mostly with respect to the case outcome, we focus on other legal proceedings. In particular, we analyze, (1) whether bond without upfront payment was granted and (2) whether, upon appeal, the charge was reduced. The choice to focus on legal proceedings with binary outcomes was motivated by their simplicity and suitability for existing fairness metrics and methodology, which typically assumes categorical, rather than continuous outputs (such as, e.g., the sentence length).

The study's goal is explicitly not the predictive modeling of the judicial system, nor does it provide a final solution for debiasing legal data. Instead, we focus on developing protocols and pipelines for identifying and understanding biases in legal data, mainly focusing on \emph{data transparency} and \emph{awareness of existing biases}. 

\subsection{Related Work}
The data-driven analysis of biases in legal data has recently gained a surge of interest. Notable work that quantitatively analyses sentencing disparities includes~\citep{justfair,mustard2001, rehavi_starr,cohen_yang,Depew2017JudgesJA}. In ~\citep{mustard2001, rehavi_starr}, the focus lies mainly on how the characteristics of the defendant influence the likelihood of certain case outcomes. In contrast,~\citep{justfair,cohen_yang,Depew2017JudgesJA} evaluate the characteristics of both the defendant and the judge. Notably,~\citep{justfair,Depew2017JudgesJA} consider identifying information on judges. The recently initiated \emph{JUSTFAIR} project~\citep{justfair} collects large-scale criminal data. It includes, besides sentencing information, also demographic information and criminal history for the defendant, as well as demographic information on the judge.

The proposed approach applies algorithmic fairness tools, which were developed in several seminal works, notably~\citep{hardt2016,dwork2012}. \citep{celis2020data} studies the problem of mitigating biases in training data in the broader machine learning context. \citep{Kleinberg30096} broadly discusses the idea of detecting discrimination and bias with algorithmic tools. The \textsc{Fairlearn} toolkit~\citep{bird2020fairlearn} implements a range of fairness metrics.

\subsection{Summary of contributions}
The goal of this study is to develop pipelines for efficiently identifying and measuring biases in large-scale legal data. Specifically, our contributions are as follows:
\begin{enumerate}
    \item In four case studies, we compare the decisions of a 	``typical'' judge with those of a ``fair'' judge that applies one of three concepts of fairness. This setup allows for quantifying biases in the underlying criminal case data and gives insight on the suitability of these notions for legal data.
    \item We analyze the impact of representation biases on fairness in legal cases by incorporating data balancing into our mitigated (``fair'') models.
\end{enumerate}
To the best of our knowledge, the present study is the first to evaluate biases in legal data from an algorithmic fairness perspective.

\section{Methods}
\noindent In this section we introduce the fairness concepts used in our study, as well as the methodology underlying both the baseline and the ``fair'' classifiers. We adapt an experimental approach introduced in~\citep{DBLP:journals/corr/abs-1803-02453} to the training of fair classifiers on legal data.  All models were implemented in \textsc{Python}.

\subsection{Notions of fairness}
\label{sec:fairness}
We begin by formally introducing the fairness concepts considered in this study. Since the study focuses on fairness aspects of judicial decisions across demographic groups, we focus on \emph{group fairness}, as opposed to individual fairness notions. Below, we give formal definitions for a prediction task with binary outcomes ($\hat{Y} \in \lbrace 0,1 \rbrace$) with respect to a binary protected characteristic ($A \in \lbrace 0,1 \rbrace$).

The first notion we consider is \emph{demographic parity}, which requires that the outcome is independent of the protected characteristic, i.e., \begin{align}
    \mathbb{P} \left[ \hat{Y}= 1 \vert A= 0 \right] = \mathbb{P} \left[ \hat{Y}=1 \vert A = 1 \right] \; .
\end{align}
Unfortunately, this notion is known to have serious shortcomings. In particular, ~\citep{dwork2012} argues that demographic parity is not sufficient to ensure fairness (especially in the presence of representation biases), and may not allow for learning the optimal predictor. To mitigate these shortcomings,~\citep{hardt2016} introduces two additional fairness metrics: Equalized odds and equalized opportunity. \emph{Equalized odds} equalizes both true positive and false positive rates across demographic groups; formally, we require
\begin{align}
    \mathbb{P} \left[ \hat{Y}=1 \vert A = 0, Y=y \right] 
    = \mathbb{P} \left[ \hat{Y} = 1 \vert A=1, Y=y \right] \quad \forall y \in \lbrace 0,1 \rbrace \; .
\end{align}
\emph{Equal opportunity} is a relaxation of equalized odds, where only the true positive rates are equalized across demographic groups, i.e.,
\begin{align}
    \mathbb{P} \left[ \hat{Y}=1 \vert A=0, Y=1 \right] = \mathbb{P} \left[ \hat{Y}=1 \vert A =1, Y=1 \right] \; .
\end{align}


\subsection{Baseline classifier}
 We consider two baseline classifiers: A \emph{weighted logistic regression model} and a \emph{gradient boosted decision tree}. Below, we briefly describe both approaches. For a more detailed overview on both approaches, see~\citep{James2013}. Our experiments use an implementation from the \textsc{scikit-learn} package\footnote{https://scikit-learn.org/stable/}. 
 
 The first approach utilizes a \emph{classical regression model} with a logistic loss function. The model returns, for a given case input, the probability of two binary outcomes: In the first experiment, whether bail without upfront payment was granted (``free bond'') and in the second, whether the charge was reduced upon appeal (``charge reduction'').
 Here, we use a \emph{weighted} approach, meaning that we incorporate weights in the training pipeline that are chosen inversely proportional to class frequencies in the input data. The class frequencies and weights are given in Appendix~\ref{apx:c2}.

 Formally, the output is computed as follows: Let $x \in \mathbb{R}^d$ denote an input feature vector (where $d$ denotes the number of features in the data, here $d=7$); $a$ and $b$ the model's coefficients and $P$ the output probability.
 \begin{align}
    P(x) = \frac{e^{a+bx}}{1+e^{a+bx}} \; .
\end{align}
Due to its simplicity, the logistic regression model is a suitable benchmark for basic classification tasks. The hyperparameters for the model are specified in Appendix~\ref{apx:c1}.
 
The second approach, a \emph{gradient boosted decision tree model}, learns an ensemble of simple decision trees (decision ``stumps''), which predict an outcome based on only one of the input features. The model combines these decision ``stumps'' iteratively into one predictor, which can then perform well on more complex classification tasks. In iteration $j$, an update of the form 
\begin{align}
    F_j(x) &= F_{j-1}(x) + \gamma_j h_j(x)\\ \gamma_j &= \argmin_\gamma \sum_{i=1}^{n} L(y_i,F_{j-1}(x_i)+\gamma h_j(x_i)) 
\end{align}
is performed, where $h_j(x)$ is a decision tree, $\gamma_j$ is a coefficient chosen to minimize a loss function $L$ (in our experiments, the logistic loss) over a training set $ \lbrace(x_i, y_i) \rbrace_{i=1}^n$. 
The gradient boosted decision tree approach implements a more complex model than the regression approach and is expected to have superior performance. We use the default hyperparameters for this model, as specified in Appendix~\ref{apx:c1}.
 

\subsection{Classifier with fairness constraints}
A first (naive) idea for training a ``fair'' classifier might be to simply ignore all protected characteristics, such as race or gender, when aiming to train an unbiased classifier. However, this approach does not mitigate biases effectively, since protected characteristics can typically be inferred from unprotected ones~\citep{pedreshi2008}. Instead, we analyze data biases using the classical algorithmic fairness notions introduced in section~\ref{sec:fairness}: 
\emph{Demographic Parity}, \emph{Equal Odds} and \emph{Equal Opportunity}. For each legal setting and crime type, we train, in addition to the baseline, three ``fair'' classifiers using the three fairness metrics.
We constrain each trained classifier to adhere to one of the fairness notions and evaluate the impact of the mitigating constraint with respect to all three fairness notions. In order to integrate the fairness constraints into our training pipeline, we follow the setup in \citep{DBLP:journals/corr/abs-1803-02453} (\emph{exponentiated-gradient reduction}\footnote{\url{https://github.com/fairlearn/fairlearn}}). 
We again consider both logistic regression and boosted decision trees.

\subsection{Data balancing}
Finally, we aim to account for \emph{representation bias}, too. We balance data by \emph{random subsampling}: To achieve an equal split between the two (binary) outcomes, we sample an appropriately sized subset from the data points with the more frequent outcome. We use \textsc{numpy}'s \emph{random shuffle}\footnote{\url{https://numpy.org/doc/stable/reference/random/generated/numpy.random.shuffle.html}} function to perform the subsampling. 

\subsection{Evaluation metrics}
We report results with respect to the following fairness metrics:
\begin{itemize}
    \item Demographic parity difference: $\Delta_{DP} := \left\vert \mathbb{P} \left[ \hat{Y}= 1 \vert A= 0 \right] - \mathbb{P} \left[ \hat{Y}=1 \vert A = 1 \right] \right\vert$
    \item True positive difference: $\Delta_{TP} := \left\vert \mathbb{P} \left[ \hat{Y}=1 \vert A=0, Y=1 \right] - \mathbb{P} \left[ \hat{Y}=1 \vert A =1, Y=1 \right] \right\vert$
    \item False positive difference: $\Delta_{FP} := \left\vert \mathbb{P} \left[ \hat{Y}=1 \vert A=0, Y=0 \right] - \mathbb{P} \left[ \hat{Y}=1 \vert A =1, Y=0 \right] \right\vert$
\end{itemize}
Naturally, a classifier with Demographic parity constraint is designed to mitigate Demographic parity difference; a classifier with Equal opportunity constraint to mitigate true positive difference and a classifier with Equalized odds constraint both true and false positive difference.

\section{Legal data}
\subsection{Data sources}

\begin{minipage}{0.5\textwidth}
\begin{table}[H]
\begin{tabular}{lccc}
      \toprule
      \thead{} & \textbf{Total} & \textbf{Narcotics} & \textbf{Theft} \\
      \midrule
      Initiations &  975,836 & 208,434 & 48,885  \\
      Dispositions & 831,582 & 167,735 & 42,129 \\
       \bottomrule
    \end{tabular}
    \caption{\textbf{Size of raw data.}\label{tab:raw-data}}
  \end{table}
\end{minipage}
\begin{minipage}{0.5\textwidth}
\begin{table}[H]
\begin{tabular}{lcc}
      \toprule
      \thead{} & \textbf{Charge reduction} & \textbf{Bond} \\
      \midrule
      Narcotics & 34,806 & 35,002  \\
      Theft & 12,862 & 12,363  \\
       \bottomrule
    \end{tabular}
    \caption{\textbf{Size of preprocessed data sets.}\label{tab:processed-data}}
  \end{table}
  \end{minipage}
 %

The data used in this study is provided by the Cook County State's Attorney's Office (Illinois) and publicly available at the following websites:
\begin{enumerate}
    \item \textbf{Initiation:}
     \url{https://datacatalog.cookcountyil.gov/Courts/Initiation/7mck-ehwz} (download on March 30, 2021 at 6:00 PM CST)
    \item \textbf{Sentencing:} \url{https://datacatalog.cookcountyil.gov/Courts/Sentencing/tg8v-tm6u} (downloaded on March 30, 2021 at 6:00 PM CST)
\end{enumerate}
Both data sets were created in February 2018. The first data set (\emph{Initiations}\footnote{An initiation is the decision to prosecute a defendant after a felony review by the State's Attorneys Office following an arrest; an indiction by a grand jury; or a direct filing by law enforcement (narcotics cases only).}) contains 38 features, the second (\emph{Dispositions}\footnote{A disposition is the completion of the fact-finding process that leads to the resolution of a case.}) 33. We focus our analysis on seven features that are common between both data sets. The choice of features is described in detail in the following section. Furthermore, we consider only narcotics and theft crimes in our analysis. The selection of crime types was guided by the statistical power of the respective data subsets. Table~\ref{tab:raw-data} lists the size of the raw data. We use a $75\%$/ $25\%$ training/testing split in the subsequent analysis.

\subsection{Preprocessing}
To generate the data sets used in the study, we filtered in terms of crime types and case outcomes: For crime types, we selected narcotics and theft cases. Furthermore, we generated data sets with binary case outcomes, with respect to charge reduction and bail. Table~\ref{tab:processed-data} lists the size of the data sets after preprocessing.

For the feature selection, we initially dropped unneeded features, such as dates, the primary charge flag (\textsc{true} for all cases), the charge count, charge disposition (guilty plea for all cases), and redundant IDs. We then performed a correlation analysis (see Figure~\ref{fig:corr}) of the remaining features and selected a subset with low pairwise correlations. 
Finally, we removed some erroneous data base entries with defendants of ages greater than 100 years old. More details on feature selection can be found in Appendix~\ref{apx:features}.

We select cases for our target crime types based on the reported updated offense category, where we combined theft and retail theft into one category. We generated the \emph{charge reduction} outcome by first selecting primary charges with plea bargains, which were accepted to receive reduced charges. We included cases for which the class of the charge was within a set of 5 degrees of severeness, where 0 is the most severe and 5 the least severe crime. By comparing the class of the charge in \emph{Initiations} with that in \emph{Dispositions}, we say that the charge was reduced, whenever the dispositions class was greater, or less severe, than the initiations class. For \emph{bail}, we consider four types of bonds: I Bond, C Bond, D Bond, and ``no bond''. No bond is the most severe, where the defendant is denied any pretrial release. I Bonds are the least severe, allowing the defendant to be released without monetary bail. C Bonds and D Bonds are similar, with both requiring payment, where C Bonds require a full payment and D Bonds require only a partial payment. We consider binary outcomes of ``no/ paid bonds'' and ``free bond''. Here we combined C Bonds, D Bonds and ``no bond'' into the ``no/ paid bonds'' outcome and call I Bonds ``free bonds''.

Our study considers four demographic groups with respect to race and gender. In order to directly apply the classical fairness constraints introduced in section~\ref{sec:fairness}, we restrict ourselves to binary protected characteristics: We consider only cases in which the defendant identifies as female or male and as either white or black. For the former, the raw data reports the gender of defendants as either female or male. We excluded cases with gender reported to be unknown. For the latter, we focus on the two groups with the highest statistical power in the data set. The restriction to white and black defendants also follows a pattern in related legal scholarship, see, e.g., ~\citep{rehavi_starr,eberhardt2019biased}, allowing for a placement of our study in the context of this body of literature. However, we note that a future, more comprehensive study should be more inclusive with respect to race and gender.

\section{Results}
\subsection{Statistical Analysis}
We start with a preliminary statistical analysis of our legal data and compare the results across the two different crime types and the four demographic groups. We also analyze arrest numbers and the age of the defendant at incident. 

\begin{wrapfigure}{r}{0.4\textwidth}
  \centering
    \includegraphics[width=\linewidth]{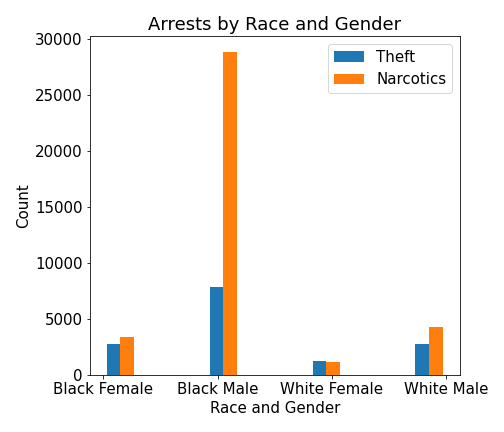}
    \caption{\textbf{Arrest counts by race and gender for both crime types}, with Theft in blue and Narcotics in Orange. The figure shows that while arrest for other demographics are relatively close between crime types, about 3 times as many black Men are arrested for Narcotics crimes. 
    }
    \label{fig:arrests}
\end{wrapfigure}

We can see in Figure~\ref{fig:arrests} that disparities are already evident in the arrest numbers across demographic groups: For both crime types, we observe that the number of arrests for black males are about three time higher than those of the other three groups. Comparing across races, we notice a larger number of black arrests; comparing across genders, we notice a larger number of male arrests. We further analyzed the number of arrests across incident cities in Cook county. The results can be found in Appendix~\ref{apx:cities}.

\begin{figure}[ht]
\begin{subfigure}{.49\textwidth}
    \centering
    \includegraphics[width=.8\linewidth]{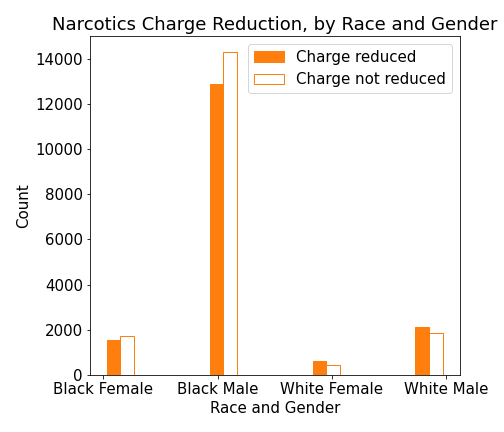}
    \label{fig:crfalse}
\end{subfigure}
\begin{subfigure}{.49\textwidth}
    \centering
    \includegraphics[width=.8\linewidth]{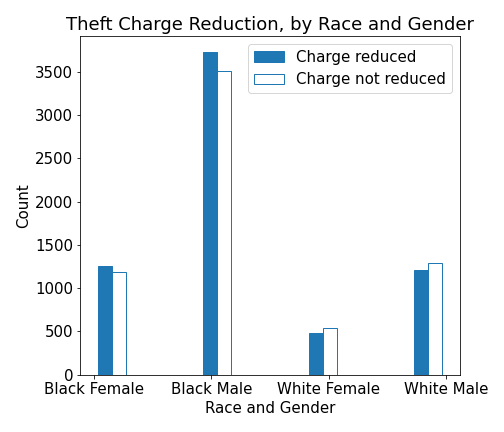}
    \label{fig:crtrue}
\end{subfigure}
\centering
\caption{\textbf{Binary outcomes for charge reduction.} Having a charge maintained means that the charge was not reduced. The percentage of arrested persons that had their charge reduced was comparable across demographic groups, except for black men in narcotics cases, who are more likely to \emph{not} have their charge reduced.}
\label{fig:cr}
\end{figure}

\begin{figure}[ht]
\begin{subfigure}{.49\textwidth}
    \centering
    \includegraphics[width=.8\linewidth]{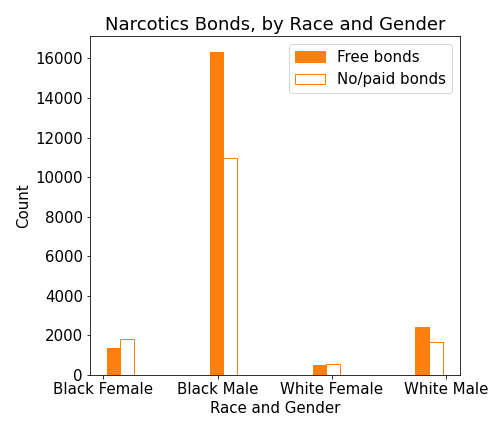}
    \label{fig:bfalse}
\end{subfigure}
\begin{subfigure}{.49\textwidth}
    \centering
    \includegraphics[width=.8\linewidth]{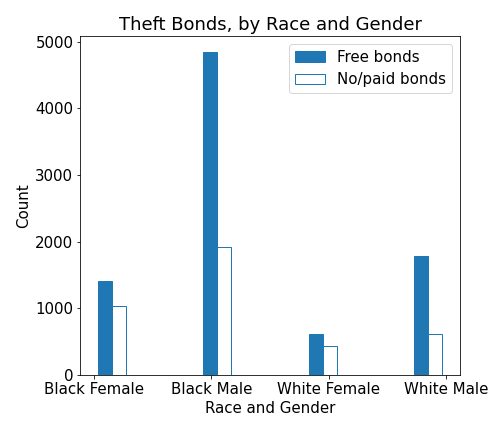}
    \label{fig:btrue}
\end{subfigure}
\centering
\caption{\textbf{Bail outcomes for Narcotics and Theft crimes.} Free bonds refer to bail outcomes that grant freedom with no payment, and no/paid bonds refer to bail outcomes that either give no freedom or require a bond payment. Generally speaking, arrested persons are more likely to be granted free bonds in cases involving theft crimes, compared to cases involving narcotics crimes. Black persons are more often given restrictive bonds, compared to white persons.}
\label{fig:b}
\end{figure}

Next, we analyze the legal decisions in our case studies (\emph{charge reduction} and \emph{free bond}, for narcotics and theft crimes) across demographic groups. We observe that 48\% of black men have their charge reduced, compared to 51\% of white men, 49\% of black women, and 52\% of white women (see also Figure~\ref{fig:cr}). 
The bail data shows that free bonds are given more often for theft crimes (68\%), compared to narcotics crime (58\%). Furthermore, black defendants are more likely to have restrictive bonds than other demographic groups, with 62\% of black men receiving a free bond, compared to 64\% of white men; and 49\% of black women, compared to 53\% of white women (Figure~\ref{fig:b}).
We note that the overrepresentation of men (especially black men) in the data sets may significantly influence the observed trends for charge reduction and free bonds.

\begin{figure}[H]
\begin{subfigure}{.49\textwidth}
\centering
\includegraphics[width=.8\linewidth]{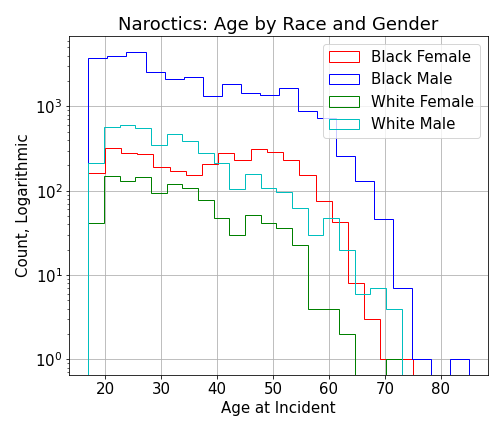}
\label{fig:narc_age}
\end{subfigure}
\begin{subfigure}{.49\textwidth}
\centering
\includegraphics[width=.8\linewidth]{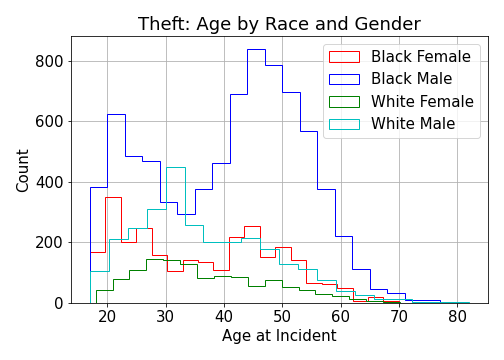}
\label{fig:theft_age}
\end{subfigure}
\caption{\textbf{Age at arrest for narcotics (left) and theft (right).} The Narcotics graph is scaled logarithmic due to the much larger proportion of black Men present in the data. Narcotics crimes are mostly young people, while Theft appears to have two peaks among arrested blacks at around 20 and 45 years old. For arrested whites, the frequency peaks around 30 years old and decreases with age.}
\label{fig:age}
\end{figure}

Other interesting trends can be found by looking at the reported age of those arrested. Figure~\ref{fig:age} shows that those arrested for narcotics crimes are mostly young people between 20 and 30 years, across all demographic groups. For theft the trend is different: White people arrested for theft crimes are mostly around 30 years old, while the age at arrest for black people peaks at around 20 years old and again around 45 years.


\subsection{Algorithmic analysis of data biases}
After running our regression models on all four case studies (charge reduction/ free bonds and narcotics/ theft crimes), we found that on average, applying fairness constraints does succeed in mitigating biases. In addition, data balancing helps to mitigate representation biases and, on average, improves fairness metrics. We observe that our baseline classifiers, which represent the actions of a ``typical'' judge, perform the least fair. Both regression models (weighted logistic regression (Tab.~3,~5) and gradient boosted tree search (Tab.~4,~6)) achieve similar fairness results and both seem to be equally affected by balancing and fairness mitigation. Notably, data balancing improved, on average, the fairness metrics for both approaches (Tab.~3-6,~first column).
\begin{table}[ht]\centering
\newcolumntype{C}{>{\centering\arraybackslash}X}
\begin{tabular}{l*{9}{S}}
\toprule
\addlinespace 
&\multicolumn{2}{c}{Baseline} 
&\multicolumn{2}{c}{Mitigated (DP)} 
&\multicolumn{2}{c}{Mitigated (EOdd)} 
&\multicolumn{2}{c}{Mitigated (EOpp)}\\
 \cmidrule(lr){2-3}
 \cmidrule(lr){4-5}
 \cmidrule(lr){6-7}
 \cmidrule(lr){8-9}
{}&{Unbalanced}&{Balanced}&{Unbalanced}&{Balanced}&{Unbalanced}&{Balanced}&{Unbalanced}&{Balanced}
\tabularnewline
\cmidrule[\lightrulewidth](lr){1-9}\addlinespace[1ex]
\textbf{Narcotics}  \tabularnewline
$\Delta_{DP}$ & 0.068 & 0.019 & 0.002 & 0.017 & 0.005 & 0.021 & 0.036 & 0.025 \tabularnewline
$\Delta_{TP}$ & 0.014 & 0.019 & 0.091 & 0.022 & 0.068 & 0.009 & 0.045 & 0.010 \tabularnewline
$\Delta_{FP}$ & 0.120 & 0.012 & 0.059 & 0.013 & 0.051 & 0.007 & 0.086 & 0.014 \tabularnewline
\textbf{Theft} \tabularnewline
$\Delta_{DP}$ & 0.121 & 0.041 & 0.031 & 0.025 & 0.0448 & 0.006 & 0.057 & 0.045 \tabularnewline
$\Delta_{TP}$ & 0.124 & 0.034 & 0.039 & 0.020 & 0.060 & 0.021 & 0.070 & 0.025\tabularnewline
$\Delta_{FP}$ & 0.108 & 0.100 & 0.013 & 0.058 & 0.019 & 0.021 & 0.034 & 0.100 \tabularnewline
\addlinespace
\bottomrule
\end{tabular}
\captionof{table}{\textbf{Logistic Regression:} Results for predictive analysis (\textbf{Charge reduction}) for baseline in comparison with Demographic parity (DP), Equalized odds (EOdd) and Equalized Opportunity (EOpp) mitigations. Here, $\Delta_{DP}$ denotes Demographic parity difference, $\Delta_{TP}$ the True positive difference and $\Delta_{FP}$ the False positive Difference.\label{tab:3}}
\end{table}

\begin{table}[ht] \centering
\newcolumntype{C}{>{\centering\arraybackslash}X}
\begin{tabular}{l*{9}{S}}
\toprule
\addlinespace 
&\multicolumn{2}{c}{Baseline} 
&\multicolumn{2}{c}{Mitigated (DP)} 
&\multicolumn{2}{c}{Mitigated (EOdd)} 
&\multicolumn{2}{c}{Mitigated (EOpp)}\\
 \cmidrule(lr){2-3}
 \cmidrule(lr){4-5}
 \cmidrule(lr){6-7}
 \cmidrule(lr){8-9}
{}&{Unbalanced}&{Balanced}&{Unbalanced}&{Balanced}&{Unbalanced}&{Balanced}&{Unbalanced}&{Balanced}
\tabularnewline
\cmidrule[\lightrulewidth](lr){1-9}\addlinespace[1ex]
\textbf{Narcotics}  \tabularnewline
$\Delta_{DP}$ & 0.097 & 0.035 & 0.027 & 0.009 & 0.028 & 0.027 & 0.053 & 0.016 \tabularnewline
$\Delta_{TP}$ & 0.018 & 0.009  & 0.057 & 0.036 & 0.051 & 0.004 & 0.028 & 0.020 \tabularnewline
$\Delta_{FP}$ & 0.145 & 0.013 & 0.082 & 0.010 & 0.078 & 0.012 & 0.104 & 0.007 \tabularnewline
\textbf{Theft} \tabularnewline
$\Delta_{DP}$ &  0.141 & 0.064 & 0.024 & 0.039 & 0.014 & 0.029 & 0.054 & 0.062 \tabularnewline
$\Delta_{TP}$ &  0.135 & 0.023 & 0.018 & 0.003 & 0.005 & 0.006 & 0.059 & 0.019 \tabularnewline
$\Delta_{FP}$ &  0.136 & 0.134 & 0.021 & 0.068 & 0.012 & 0.050 & 0.039 & 0.127 \tabularnewline
\addlinespace
\bottomrule
\end{tabular}
\captionof{table}{\textbf{Gradient Boosting:} Results for predictive analysis (\textbf{Charge reduction}) for baseline in comparison with Demographic parity (DP), Equalized odds (EOdd) and Equalized Opportunity (EOpp) mitigations. Here, again, $\Delta_{DP}$ denotes Demographic parity difference, $\Delta_{TP}$ the True positive difference and $\Delta_{FP}$ the False positive Difference.\label{tab:4}}
\end{table}
As expected, applying a specific fairness constraint improves its target metric.
We observe that applying a demographic parity constraint decreases the demographic parity difference, which indicates that our baseline was not classifying independent of race and gender (Tab.~3-6,~second column). We also see that equalized odds mitigation decreases both true positive and the false positive difference (Tab.~3-6, third column), and that equalized opportunity decreases the true positive difference, meaning
that the accuracy of our classifier varied across demographic groups (Tab.~3-6, fourth column). 
We further observe that equalized opportunity mitigation often decreases the true positive difference more than equalized odds, while either maintaining or increasing the false positive difference. This suggests that optimizing one fairness metric might worsen others. 
Alongside our mitigation techniques, we used data balancing to counteract representation bias. When applied to the baseline classifier, data balancing generally improves the fairness metrics. 
\begin{table}[ht] \centering
\newcolumntype{C}{>{\centering\arraybackslash}X}
\begin{tabular}{l*{9}{S}}
\toprule
\addlinespace 
&\multicolumn{2}{c}{Baseline} 
&\multicolumn{2}{c}{Mitigated (DP)} 
&\multicolumn{2}{c}{Mitigated (EOdd)} 
&\multicolumn{2}{c}{Mitigated (EOpp)}\\
 \cmidrule(lr){2-3}
 \cmidrule(lr){4-5}
 \cmidrule(lr){6-7}
 \cmidrule(lr){8-9}
{}&{Unbalanced}&{Balanced}&{Unbalanced}&{Balanced}&{Unbalanced}&{Balanced}&{Unbalanced}&{Balanced}
\tabularnewline
\cmidrule[\lightrulewidth](lr){1-9}\addlinespace[1ex]
\textbf{Narcotics}  \tabularnewline
$\Delta_{DP}$ &  0.062 & 0.036 & 0.010 & 0.026 & 0.006 & 0.0001 & 0.012 & 0.027 \tabularnewline
$\Delta_{TP}$ &  0.067 & 0.011 & 0.016 & 0.011 & 0.016 & 0.007 & 0.010 & 0.002 \tabularnewline
$\Delta_{FP}$ &  0.065 & 0.083 & 0.012 & 0.065 &  0.004 & 0.020 & 0.025 & 0.072 \tabularnewline
\textbf{Theft} \tabularnewline
$\Delta_{DP}$ &  0.103 & 0.027 & 0.004 & 0.009 & 0.024 & 0.004 & 0.007 & 0.025 \tabularnewline
$\Delta_{TP}$ &  0.108 & 0.029 & 0.012 & 0.017 & 0.042 & 0.013 & 0.006 & 0.033 \tabularnewline
$\Delta_{FP}$ &  0.079 & 0.077 & 0.022 & 0.030 & 0.010 & 0.001 & 0.023 & 0.077 \tabularnewline
\addlinespace
\bottomrule
\end{tabular}
\captionof{table}{\textbf{Logistic Regression:} Results for predictive analysis (\textbf{Bond}) for baseline in comparison with Demographic parity (DP), Equalized odds (EOdd) and Equalized Opportunity (EOpp) mitigations. Here, again, $\Delta_{DP}$ denotes Demographic parity difference, $\Delta_{TP}$ the True positive difference and $\Delta_{FP}$ the False positive Difference.\label{tab:5}}
\end{table}

\begin{table}[ht] \centering
\newcolumntype{C}{>{\centering\arraybackslash}X}
\begin{tabular}{l*{9}{S}}
\toprule
\addlinespace 
&\multicolumn{2}{c}{Baseline} 
&\multicolumn{2}{c}{Mitigated (DP)} 
&\multicolumn{2}{c}{Mitigated (EOdd)} 
&\multicolumn{2}{c}{Mitigated (EOpp)}\\
 \cmidrule(lr){2-3}
 \cmidrule(lr){4-5}
 \cmidrule(lr){6-7}
 \cmidrule(lr){8-9}
{}&{Unbalanced}&{Balanced}&{Unbalanced}&{Balanced}
&{Unbalanced}&{Balanced}&{Unbalanced}&{Balanced}
\tabularnewline
\cmidrule[\lightrulewidth](lr){1-9}\addlinespace[1ex]
\textbf{Narcotics}  \tabularnewline
$\Delta_{DP}$ &  0.076 & 0.074 & 0.004 & 0.013 & 0.005 & 0.029 & 0.052 & 0.085 \tabularnewline
$\Delta_{TP}$ &  0.044 & 0.004 & 0.016 & 0.003 & 0.002 & 0.016 & 0.024 & 0.001 \tabularnewline
$\Delta_{FP}$ &  0.124 & 0.164 & 0.019 & 0.001 & 0.007 & 0.020 & 0.096 & 0.181 \tabularnewline
\textbf{Theft} \tabularnewline
$\Delta_{DP}$ &  0.026 & 0.051 & 0.023 & 0.022 & 0.020 & 0.031 & 0.026 & 0.048 \tabularnewline
$\Delta_{TP}$ &  0.009 & 0.009 & 0.009 & 0.009 & 0.005 & 0.005 & 0.009 & 0.025 \tabularnewline
$\Delta_{FP}$ &  0.053 & 0.104 & 0.047 & 0.046 & 0.046 & 0.051 & 0.053 & 0.115 \tabularnewline
\addlinespace
\bottomrule
\end{tabular}
\captionof{table}{\textbf{Gradient Boosting:} Results for predictive analysis (\textbf{Bond}) for baseline in comparison with Demographic parity (DP), Equalized odds (EOdd) and Equalized Opportunity (EOpp) mitigations. Here, again, $\Delta_{DP}$ denotes Demographic parity difference, $\Delta_{TP}$ the True positive difference and $\Delta_{FP}$ the False positive Difference.\label{tab:6}}
\end{table}

Overall, \emph{equalized odds with balancing} provided the best mitigation in our case studies. This suggests that our current judicial model could benefit from sentencing guidelines that equalize the rate of
positive and negative outcomes across demographic groups.
One of the primary objectives of sentencing guidelines is to reduce the sanctioning of innocent defendants, i.e., to reduce false positive rates in sentencing. Therefore, imposing fairness constraints that equalize false positive rates across demographic groups is a promising avenue for mitigating sentencing disparities. Furthermore, our results suggest that the overreperesentation of black men in criminal case data contributes significantly to the observed biases. Mitigation approaches that recover the true distribution across demographic groups (e.g., in case precedents) could help to counteract such biases. 

We again emphasize that this study does not aim on modeling the judicial decision process, but rather on identifying and understanding biases in legal data. Therefore, we do not evaluate the accuracy of our ``typical'' and ``fair'' judges with respect to the ground truth and focus solely on analyzing fairness metrics. 

%

\section{Discussion}
Guided by the question of \emph{efficiently identifying and measuring biases in large-scale legal data}, we introduced a pipeline for the data-driven analysis of representation biases and sentencing disparities across demographic groups. We implemented two classical regression approaches and three commonly used fairness metrics to evaluate biases in criminal case data provided by the Cook County State's Attorney's Office. In particular, we presented four case studies, considering two different crime types (narcotics and theft), as well as two different judicial decisions (charge reduction and free bonds). When comparing our baseline model, which represents the decisions of a ``typical'' judge according to the data, with that of a ``fair'' judge that applies one of three concepts of fairness, we observe a quantifiable difference in our fairness metrics. 

The present study is, to our knowledge, the first to evaluate biases in legal data from an algorithmic fairness perspective. 
The results of our study echo general findings in previous quantitative studies~\cite{rehavi_starr,cohen_yang,Schanzenbach2007ReviewingTS}, albeit focusing not on sentencing disparities in the final case outcome, but on other aspects of the legal process. Looking ahead, we hope that this study sheds some light on suitable notions of fairness for legal proceedings and on  pipelines for systematically analyzing bias in large-scale legal data. 

While the present study does not evaluate the accuracy of the judicial decisions of the ``fair'' and the ``typical'' judges, future work could look at the implications of fairness constraints on metrics such as the likelihood of repeated offenses (for charge reduction) or failure to appear in court (free bonds). Note that both approaches (the ``fair'' and the ``typical'' judge) aim to model average judicial decisions and do not account for individual differences between judges. For instance, the approach is not able to measure the impact of a few ``unfair" judges in a larger group of mostly ``fair" judges. We leave the study of such cases to future work.

Other directions for future work include a study of sentencing decisions, focusing on disparities in the final case outcome across demographic groups. This requires the adaption of the 
binary fairness metrics considered here to real-valued outputs. Furthermore, while the present study focused on fairness across demographic groups (\emph{group fairness}), an interesting direction for future work would be the study of individual fairness notions. Finally, while much of the literature on biases in legal data has focused on criminal cases, we hope to study biases in civil cases too.



\newpage
\bibliographystyle{ACM-Reference-Format}
\bibliography{refs}

\appendix
\newpage
\section{Feature selection}\label{apx:features}
\begin{figure}[t]
    \centering
    \includegraphics[width=\linewidth]{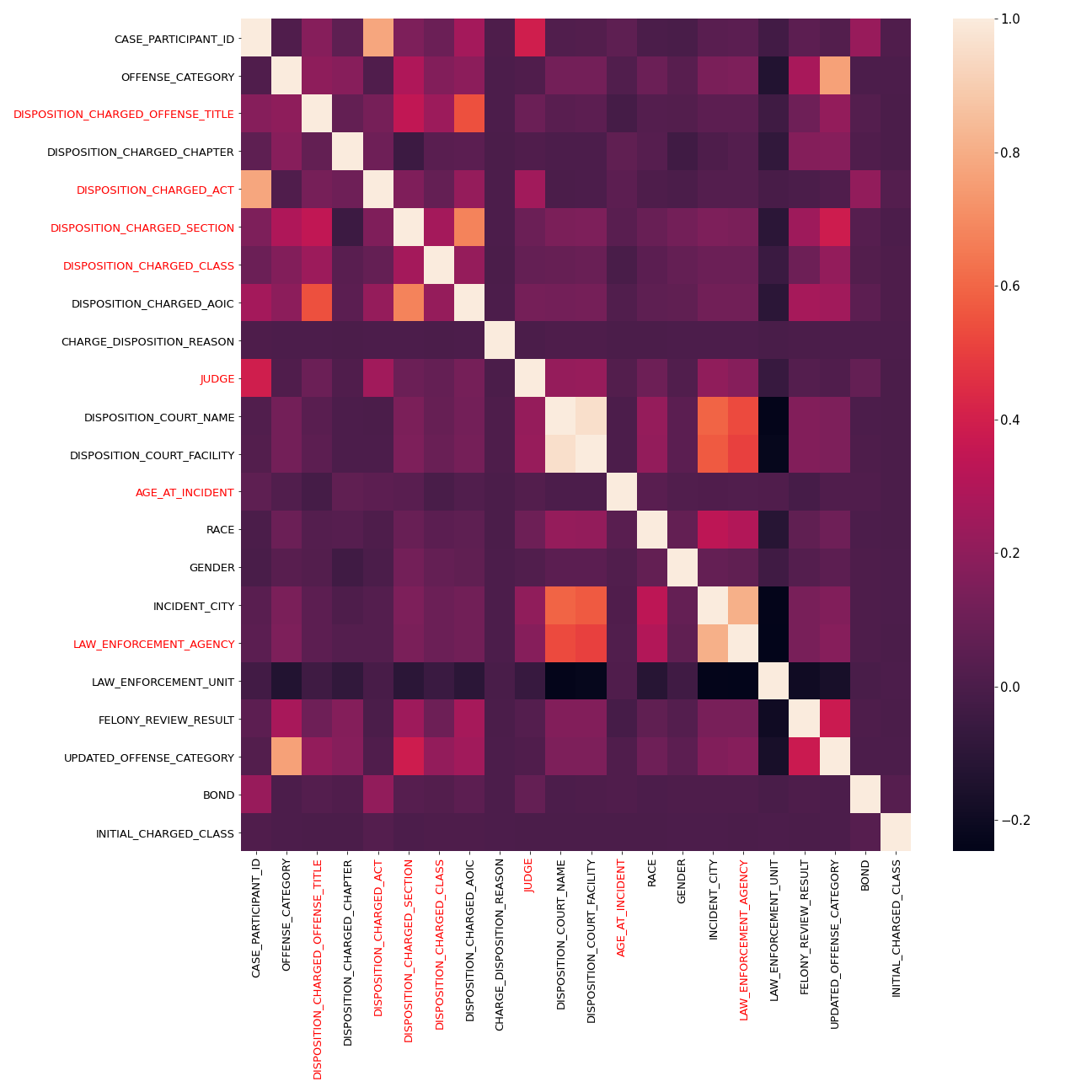}
    \caption{Feature correlation in the post processed data. The features highlighted in red are used in our predictions. Lighter colors indicate a greater correlation.}
    \label{fig:corr}
\end{figure}
Figure~\ref{fig:corr} shows the correlation between all features\footnote{A glossary of all features with detailed descriptions can be found at \url{https://datacatalog.cookcountyil.gov/api/views/7mck-ehwz/files/cec52aad-1f1a-4bf3-b02d-e5360298f66e?download=true&filename=CCSAO\%20Data\%20Glossary.pdf}.} in the preprocessed data. The following features were selected for the analysis: 
\begin{itemize}
    \item \textbf{Act, Section:} Legal act/ section for the charge according to the Illinois criminal statute. Both features provide information on the type of the crime and the context of the crime situation.
    \item \textbf{Age\_At\_Incident:} Age of defendant at the date of the incident as reported in the case intake.
    \item \textbf{Class:} Legal class for the charge.
    \item \textbf{Judge:} Judge who oversaw the case.
    \item \textbf{Law\_Enforcement\_Agency:} Law Enforcement agency associated with the arrest.
    \item \textbf{Disposition\_Charged\_Offense\_Title:} Specific title of the charged offense at disposition.  
\end{itemize}

\section{Incident cities}
\label{apx:cities}
\begin{figure}[t]
    \centering
    \includegraphics[width=0.5\linewidth]{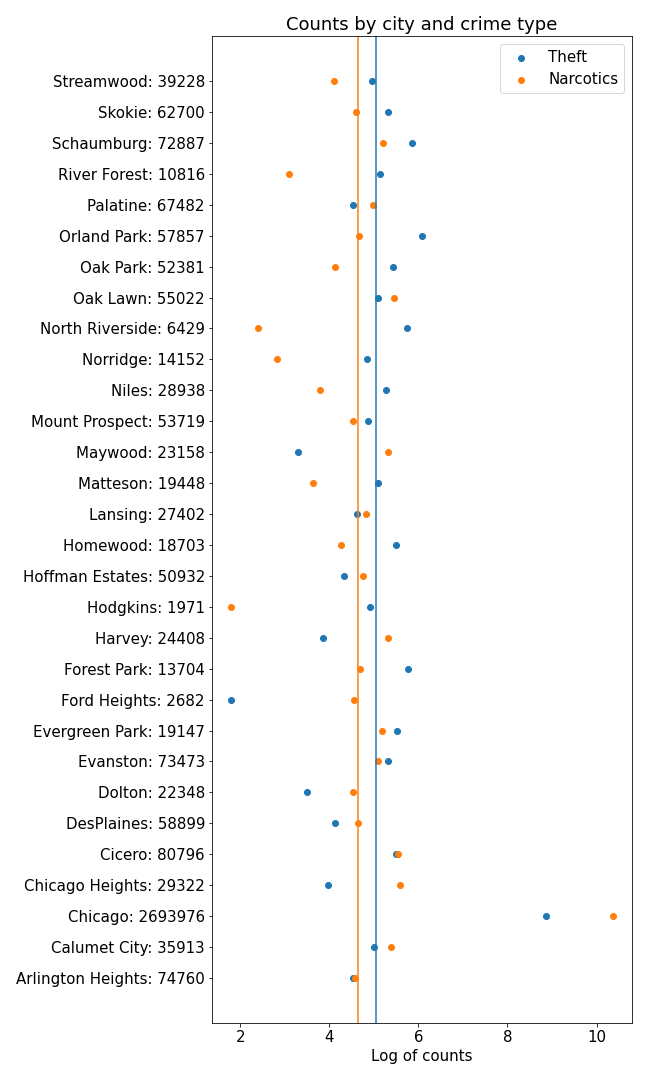}
    \caption{Arrest counts across cities in Cook County by offense category. Cities are given with their estimated total population as of 2019, and arrest counts are given on a logarithmic scale. The lines represent the median count for each crime type.}
    \label{fig:cities}
\end{figure}
Figure~\ref{fig:cities} shows arrest counts in 30 out of 142 incident cities in Cook County in the raw data. The shown data accounts for $89\%$ of all arrests in Narcotics and Theft in Cook County included in the data set.

\section{Modeling Information}
\subsection{Hyperparameters}\label{apx:c1}

\begin{table}[ht] \centering
\newcolumntype{C}{>{\centering\arraybackslash}X}
\begin{tabular}{c c}
\toprule
\addlinespace 
{Model}&{Parameters}
\tabularnewline
\cmidrule[\lightrulewidth](lr){1-2}\addlinespace[1ex]

\textbf{Logistic Regression} & max\_iter $=$ 5000
\tabularnewline
& class\_weight $=$ 'balanced' \tabularnewline

\textbf{Gradient Boosting} & Default \tabularnewline
\addlinespace
\bottomrule
\end{tabular}
\captionof{table}{The hyperparameters used to train our two types of models. All Gradient Boosting or unspecified Logistic Regression parameters are left as \textsc{sklearn}'s default.\label{tab:7}}
\end{table}

\subsection{Class Frequencies and Weights}\label{apx:c2}

\begin{table}[ht] \centering
\newcolumntype{C}{>{\centering\arraybackslash}X}
\begin{tabular}{l*{9}{S}}
\toprule
\addlinespace 
&\multicolumn{2}{c}{Frequency} 
&\multicolumn{2}{c}{Weight} \\
 \cmidrule(lr){2-3}
 \cmidrule(lr){4-5}
{}&{True}&{False}&{True}&{False}
\tabularnewline
\cmidrule[\lightrulewidth](lr){1-5}\addlinespace[1ex]
\textbf{Narcotics}  \tabularnewline
Charge Reduction & 0.484 & 0.516 & 1.033 & 0.969 \tabularnewline
Bond & 0.576 & 0.424 & 0.867 & 1.180 \tabularnewline

\textbf{Theft} \tabularnewline
Charge Reduction &  0.508 & 0.492 & 0.984 & 1.017 \tabularnewline
Bond &  0.683 & 0.317 & 0.732 & 1.580  \tabularnewline
\addlinespace
\bottomrule
\end{tabular}
\captionof{table}{Class frequencies and weights across sentencing features and crime types. The weights are those used in our weighted logistic regression model, and calculation specifics can be found at \nolinkurl{https://scikit-learn.org/stable/modules/generated/sklearn.linear_model.LogisticRegression.html}\label{tab:8}}
\end{table}

\end{document}